# Imparting icephobicity with substrate flexibility

*Thomas Vasileiou, Thomas M. Schutzius\*, Dimos Poulikakos\*.*

Laboratory of Thermodynamics in Emerging Technologies, Department of Mechanical and

Process Engineering, ETH Zurich, Sonneggstrasse 3, CH-8092 Zurich, Switzerland.

ABSTRACT. Ice accumulation hinders the performance of, and poses safety threats for infrastructure both on the ground and in the air. Previously, rationally designed superhydrophobic surfaces have demonstrated some potential as a passive means to mitigate ice accretion; however, further studies on material solutions that reduce impalement and contact time for impacting supercooled droplets (high viscosity) and can also repel droplets that freeze during surface contact are urgently needed. Here we demonstrate the collaborative effect of substrate flexibility and surface micro/nanotexture on enhancing both icephobicity and the repellency of viscous droplets (typical of supercooled water). We first investigate the influence of increased viscosity (spanning from 0.9 to 1078 mPa·s using water-glycerol mixtures) on impalement resistance and droplet-substrate contact time after impact. Then we examine the effect of droplet partial solidification on recoil and simulate more challenging icing conditions by impacting supercooled water droplets (down to −15 °C) onto flexible and rigid surfaces containing ice nucleation promoters (AgI). We demonstrate a passive mechanism for shedding partially solidified (recalescent) droplets–under conditions where partial solidification occurs much faster than the natural droplet oscillation–which does not rely on converting droplet surface energy into kinetic energy (classic recoil mechanism). Using an energy-based model (kinetic-elastic-capillary), we identify a previously unexplored mechanism whereby the substrate oscillation and velocity govern the rebound process, with low-areal density and moderately stiff substrates acting to efficiently absorb the incoming droplet kinetic energy and rectify it back, allowing droplets to overcome adhesion and gravitational forces, and recoil. This mechanism applies for a range of droplet viscosities, spanning from low to high viscosity fluids and even ice slurries, which do not rebound from rigid superhydrophobic substrates. For low viscosity, i.e water, if the substrate oscillates faster than the droplet spreading and retraction, the action of the substrate is decoupled from the droplet oscillation, resulting in reduction in the droplet-substrate contact time.

## Introduction

Much of the modern infrastructure exposed to the open atmosphere is susceptible to surface icing, which can compromise aircraft airworthiness,[1,2] hinder wind turbine power generation,[3,4] and disable power lines.[5,6] A very common mechanism of icing in cold climates and seasons is associated with supercooled water droplets solidifying upon contact with surfaces. Such conditions are encountered by aircraft inside clouds[7] and by ground materials ranging from equipment to garment exposed to freezing rain.[8,9] Supercooled water is a metastable liquid phase existing practically always at subfreezing temperatures that may solidify spontaneously without (homogeneous nucleation) or with an external nucleation seed (heterogeneous nucleation).[10] Moreover, supercooled water undergoes a 3 to 4 fold increase in viscosity at −15 °C relative to ambient temperature.[11] Both the metastable nature and high viscosity makes repelling supercooled water rather unique and much more difficult than the ambient case.[12-14]

Due to their excellent performance in terms of droplet repellency under ambient conditions, superhydrophobic surfaces have gained attention as a passive technique to mitigate ice accretion.[15-17] Generally, such surfaces consist of an array of micro-textured features decorated with nano-texture that are coated with a conformal hydrophobic layer. They are characterized by high droplet mobility due to the presence of an intervening air layer. Further, they can aid icephobicity through rapidly repelling supercooled droplets,[12,18-20] minimization of droplet-substrate heat transfer,[21,22] nucleation delay enhancement,[23,24] ice adhesion reduction,[25,26] and facile defrosting.[18,27] For droplet removal, icephobicity depends on maintaining the intervening air displace the intervening air layer, resulting in the so-called Cassie–Baxter-to-Wenzel wetting transition.[28,29] In this case, the droplet sticks on the surface resulting eventually in freezing. The ability of the texture to prevent the Cassie–Baxter-to-Wenzel transition is associated to its impalement resistance and quantified by the threshold impact velocity at which the transition takes place.

Beyond the air layer displacement, droplet mobility loss on superhydrophobic surfaces can also occur when the liquid viscosity is too high.[30,31] To underpin this statement, droplet mobility loss due to high viscosity has even been reported for sublimating surfaces, where impalement is not an issue.[32] Likewise, we expect droplet mobility loss to occur at the onset of rapid recalescent freezing, whereby a solid-liquid slurry mixture is generated[10] and the excess droplet surface energy cannot be efficiently rectified back into kinetic energy.[18] However, to date solidification during impact of supercooled droplet under isothermal condition in the presence of realistic ice seeds has not been sufficiently explored; droplet icing has been achieved either because of significant substrate cooling compared to the droplet[13, 18] (substrate temperatures lower than −25 °C) or after prolonged exposure to supercooled water streams.[24, 33] For real-life applications, in spite of the rationally designed surface texture and chemistry that can be used to inhibit nucleation, contamination of the surface may provoke nucleation even at low-levels of supercooling.

Recently, substrate flexibility—which is a property of many natural and synthetic hydrophobic materials—was shown to have a collaborative effect with existing hydrophobic micro/nanotexture on enhancing superhydrophobicity as defined by a rise in the impalement resistance[34] and a reduction in the droplet-surface contact time.[34,35] Yet, for flexible materials, it is unclear how viscosity and solidification affect droplet-substrate impact behavior. Such fundamental understanding is important for the rational design of icephobic surfaces. Here we show that substrate flexibility and hierarchical hydrophobic surface texture can work collaboratively towards enhancing the mobility[10] of impacting liquid droplets (boosting icephobicity and amphiphobicity[30]), as defined by contact time ($t_c$) and the impalement resistance, with a range of dynamic viscosities ($\mu = 0.9$ to 1078 mPa·s) and degrees of solidification ($\phi = 0$ to 0.2), many of which are impossible for rigid materials. Droplet-substrate contact time reduction, so-called "pancake bouncing",[36] is demonstrated for a range of liquid viscosities, which can render other techniques ineffective.[37] To interpret the above behavior and facilitate the development of icephobicity and amphiphobicity design rules, we develop models based upon energy conservation considerations, for predicting viscous and ice slurry droplet rebound and contact time reduction from flexible substrates based on the substrate mass and stiffness. Furthermore, we investigate the effect of contamination (nucleation promoting seeds) on supercooled droplet impact behavior and define a threshold nucleation rate based upon droplet supercooling, surface composition, and degree of contamination, above which the freezing dynamics are faster than the recoil dynamics. We then show that in spite of the instantaneous freezing that occurs, the potential energy stored in the elastic substrate is able to be rectified into outgoing droplet kinetic energy and the partially solidified droplet can be ejected from the surface overcoming adhesion, viscous, and gravitational forces. This represents an unexplored and important corollary mechanism to icephobicity, the self-cleaning of nucleating particles from surfaces with solidifying water.

## Experimental Section

*Materials*

We obtained the following chemicals from Sigma Aldrich: poly(methyl methacrylate) powder (PMMA, crystalline, $M_w$ ~996000 Da), poly(vinylidene fluoride) pellets (PVDF, $M_w$ ~ 71000 Da), surface modified nanoclay (0.5 – 5 wt. % aminopropyltriethoxysilane, 15 – 35 wt. % octadecylamine), silica nanopowder ($SiO_2$, primary particle size 12 nm), silver iodide (AgI, 99 %), N-methyl-2-pyrrolidone (NMP, 99.5 wt. %), acetic acid (≥ 99.7 wt. %), hydrochloric acid (37 wt. %) and glycerol (≥ 99.5 wt. %). We acquired the fluoroacrylic copolymer (PMC, 20 wt. % in water, Capstone ST-100) from DuPont. We obtained the hydrophobic fumed silica (HFS, Aerosil

R 8200) from Evonik, acetone (≥ 99.5 wt. %) and isopropyl alcohol (IPA, ≥ 99.5 wt. %) from Thommen-Furler AG and hydrogen peroxide (30 wt. %) from VWR. We purchased low-density polyethylene (LDPE) food packing film with a thickness of 12.5 μm from a local supplier.

*Dispersion preparation*

We used two hydrophobic nanocomposite coatings in this work, one with low impalement resistance (nC1) and one with high impalement resistance (nC2). The coatings consisted of a fluorinated polymer, in order to achieve low surface energy, and nanoparticles to increase the surface roughness. A detailed list of the ingredients and their concentrations is given in Table 1.

For the coating nC1, we prepared stock solutions of 10 wt. % PVDF in NMP and 10 wt. % of PMMA in acetone separately by dissolving the polymers under slow mechanical mixing overnight at 50 °C and at room temperature, respectively. We mixed 400 mg of nanoclay platelets and 100 mg of HFS in 8000 mg acetone in a 10 mL vial and we treated the mixture with a probe sonication (130 W, 3 mm probe, 60 % amplitude, 20 kHz frequency, Sonics Vibracell, VCX-130) for 30 sec. Once a stable suspension was formed, we added 440 mg of 10 wt. % PVDF and 250 mg of 10 wt. % PMMA and we stirred mechanically at room temperature for about 1 min.

For the coating nC2,[*,**] we suspended 140 mg SiO$_2$ particles in a mixture of 600 mg of acetic acid and 4000 mg acetone in a 10 mL vial and we sonicated the mixture for 6 min. In a separate 10 mL vial, we mixed 500 mg acetic acid with 4000 mg of acetone and subsequently, we added dropwise 730 mg of 20 wt. % PMC while stirring magnetically at 650 rpm. The nanoparticle suspension was added to the polymer solution while stirring magnetically at 1000 rpm.

**Table 1.** Composition of hydrophobic coatings

| Ingredient | nC1 Concentration, wt. % | nC2 Concertation, wt. % |
|---|---|---|
| PVDF | 0.5 | 0.0 |
| PMMA | 0.3 | 0.0 |
| PMC | 0.0 | 1.4 |
| Nanoclay | 4.4 | 0.0 |
| HFS | 1.1 | 0.0 |
| SiO2 | 0.0 | 1.5 |
| NMP | 4.3 | 0.0 |
| Acetone | 89.5 | 80.2 |
| Acetic acid | 0.0 | 11.0 |
| Water | 0.0 | 5.9 |

*Surface characterization*

Flexible superhydrophobic surfaces were generated by spray coating polymer-nanoparticle dispersions onto low-density polyethylene (LDPE) films (thickness: 12.5 μm). We deposited the dispersion on the films with an airbrush (Paasche VL, 0.73 mm head) using compressed air at 2 bar at a fixed distance of ~10 cm. We dried the samples in the oven at 90 °C for 30 min to remove any remaining solvent.

To characterize the coating wettability, we used a contact angle goniometer (OCA 35, Dataphysics Instruments) to measure the advancing ($\theta_a$) and receding ($\theta_r$) contact angles and the droplet sliding angle ($\alpha$) by the tilting cradle method. For each measurement, we gently deposited a 6 μL droplet on the coating, which had been placed on a motorized tilting stage with graduation accuracy of 0.1 deg. The stage was inclined with 0.3 deg·s$^{-1}$ while a digital camera acquired images of the droplet. Dedicated software (SCA202, Dataphysics Instruments) extracted $\theta_a$ and $\theta_r$ values from the images, in the front and the back of the droplet respectively, just before the droplet started to move. We defined $\alpha$ as the minimum inclination angle at which the droplet started sliding. We repeated each experiment at five different spots of each coating using water, glycerol, and water-glycerol mixtures. Additionally, the same setup and software can be used for calculating the surface energy of liquids using the pendant droplet technique.

Furthermore, we characterized the coating morphology using a dark field optical microscope (BX 60, Olympus) equipped with a digital camera (SC 50, Olympus). The measured values for $\theta_a$, $\theta_r$, and $\alpha$ along with representative micrographs for the two coatings are shown in Figure S1. The wetting characteristics for nC1 and nC2 are similar and almost independent of the presence of glycerol.

*Droplet Impact – Ambient temperature*

We investigated the effect of viscosity $\mu$, ranging from 0.9 to 1078 mPa·s, on droplet-substrate impact dynamics by using water and different concentrations of water-glycerol mixtures at ambient conditions (~23 °C and ~45 % relative humidity). We produced droplets of almost constant diameter ($D_0 = 2.1 - 2.4$ mm) from calibrated needles with repeatability error less than 6 %, and released them from different heights in order to vary the impact velocity ($U_0$; direction normal to the surface) and the Weber number, $We = \rho D_0 U_0^2 / \sigma$, where $\rho$ and $\sigma$ are the fluid density and surface tension respectively. We use the Ohnesorge number, $Oh = \mu / \sqrt{\rho \sigma D_0} = \sqrt{We}/Re$, to relate fluid viscosity to its inertia and surface tension, where $Re = \rho D_0 U_0 / \mu$ is the Reynolds number. The water-glycerol mixture properties have been previously reported for a wide range of temperatures and concentrations.[40-41]

For all the experiments, we sectioned the flexible substrates (30 mm × 12 mm) and suspended them by mounting their shorter edge on posts. One post is stationary whereas the other is mounted on a linear stage and can be moved. This way the strain applied on the film can be precisely controlled (see Figure S2a). For the reference case, we attached the samples on a glass slide with a double adhesive tape, producing a rigid substrate. The impact events were recorded by a high-speed camera (Phantom V9.1, Vision Research) with a frame rate of 4600 s[1] using a back-illuminated configuration. A detailed description of the experimental setup can be found elsewhere.[34]

We assessed two different aspects of altering droplet mobility through substrate flexibility: the impalement resistance and the contact time $t_c$. In the first case, we impacted droplets with varying values of $We$ onto the surfaces, starting from $We \approx 2$ and increasing until we observed droplet mobility loss. We used the low impalement resistance coating (nC1), so as to avoid the effects associated with elevated droplet impact velocities (splash). We mounted the flexible substrates without strain, allowing for minimum stiffness and a marked effect of flexibility on impalement resistance.[34] We recorded three possible wetting states from each impact event: a) total rebound, when the droplet recoiled from the surface leaving no liquid behind; b) partial rebound, when part of the droplet recoiled from the surface while the other portion remained attached; or c) no rebound, when the droplet was unable to recoil from the surface. For hierarchical random surface textures, like the ones used in this study, the manifestation of the different wetting states is of stochastic nature due to multiple transitioning states (micro- and nano-Cassie).[29, 42] In other words, to characterize such behavior, one should use the same impact conditions and measure the probability of observing each impact outcome, which we denote by $\Phi_r$ for total rebound, $\Phi_p$ for partial rebound and $\Phi_n$ for no rebound.[34] Consequently, for each $We$ and $Oh$ combination, we tested five different LDPE samples and impacted each sample three times at the same spot, gathering in total 15 measurements. Then, we measured the occurrence of each outcome and divided by the total number of measurements in order to estimate $\Phi_r$, $\Phi_p$ and $\Phi_n$.

To study the effect of $\mu$ and degree of solidification ($\phi$) on $t_c$, we used samples coated with a high impalement resistance coating (nC2). The setup configuration and sample dimensions were kept the same as before, but for these experiments, the flexible samples were fixed with ~0.5 % strain. This amount of strain is suitable for observing reduction in $t_c$ without stretching considerably the superhydrophobic coating.[34] We used water, glycerol, and three water-glycerol mixtures, producing a range of $Oh$ that spanned three orders of magnitude (from $2.2 \times 10^{-3}$ to 2.6). It has been proposed that $t_c$ reduction takes place only if the $We$ is above a threshold value,[35] which we denote by $We_c$. In order to determine $We_c$ for a given $Oh$, we started by selecting two values of $We$ at the extreme ends of our search range. One value was sufficiently low so that normal rebound always occurred, and one was sufficiently high so that $t_c$ reduction was always observed. Then, we used an iterative bisection method,[43] where at each iteration we test the intermediate $We$, and we updated the search space accordingly. For each value of $We$, we repeated the droplet-substrate impact experiment five times. We terminated the procedure either if the search space was confined enough or if we acquired inconsistent results on the five repetitions.

*Droplet Impact – Subfreezing temperature*

We performed droplet-substrate impact experiments at subfreezing environmental temperatures. The environment, substrate, and droplet temperature were controlled with an insulated chamber cooled by a constant supply of dry vapor produced by boiling liquid nitrogen (see Figure S2b). We investigated the effect of flexibility on $t_c$ for temperatures ranging from 15 °C to −15 °C using the same coatings and sample support as in the ambient temperature experiments. We note at this point that imposing an accurate and uniform strain on the flexible substrate was not possible in this setup due to space restriction. The droplet impact experiments were performed under quasi-isothermal conditions. The temperature difference between the substrate and the droplet was kept below 4 °C. The reported droplet temperatures ($T_d$) have been corrected to account for evaporation due to the dry nitrogen environment; the calibration for the correction and a detailed description of the insulated chamber can be found at Ref.[12]. The properties of supercooled water have been documented in previous studies.[11,44,45]

Finally, we impacted supercooled water droplets on artificially contaminated by design flexible and rigid substrates. We impacted each droplet onto a different spot on the surface, in order to ensure that the level of contamination was kept the same. All impacts were recorded with the same high-speed camera and configuration as in the ambient temperature experiments.

*Artificial contamination*

As a contaminant, we selected AgI particles, which are a known ice nucleation promoter.[46] To control the degree of surface contamination, we used the following procedure. For low-level contamination, we cleaned glass microscopy coverslips by treating them for 7 min with a piranha solution (1:2 solution of hydrogen peroxide in hydrochloric acid), followed by bath sonication for 1 min in acetone and 1 min in isopropyl alcohol. Finally, we moved and stored the sample in deionized water until use. We prepared dispersions of AgI in acetone with different concentrations. The dispersions were treated with probe sonication for 1 min. We placed sections of the LDPE film coated with the nC2 coating on microscopy glass slides alongside the clean coverslips, after drying them under nitrogen flow. We sprayed the AgI dispersion on both the sample and the coverslip using an airbrush while keeping the substrate temperature at 50 °C. Since AgI does not form a stable dispersion in acetone and precipitates relatively fast, we took special care to mechanically stir the mixture just before the spaying procedure. Micrographs of the coverslip were acquired for five random spots before and after the artificial contamination in order to assess the degree of surface coverage, $\psi_c$, by the AgI particles. We define $\psi_c$ as the difference between the portion of the area covered by particles before and after the contamination. For high-level contamination, we sprinkled the AgI particles on the sample using a sieve until the surface turned yellowish and tilted the sample more than 90° to eliminate possible pileups. In these cases, we assumed that every part of the droplet surface touching the substrate has a practically 100 % probability to come in contact with some nucleating particle and therefore $\psi_c \approx 100$ %. Still, the actual coverage may be a little smaller since the substrate was not completely covered with AgI particles.

*Statistical and error analysis*

For all values reported, we estimated the mean value and the 95 % confidence interval of the mean (CI) from the collected measurements. Moreover, we calculated the theoretical propagated measurement error, given the accuracy of our measuring equipment. We assume that if the CI is greater than the propagated measurement error, then the latter is contained in the CI. In the following, we report the either the CI interval or the propagated measurement error, whichever is greater, trying to capture both effects emerging from the independent experiments and the measurement equipment. The sample size ($n$, number of independent experiments) and the total number of experiments ($n'$, sum of replications for all independent experiments) are reported in the legend of each figure, wherever applicable. Finally, we report the 95 % confidence interval of the population, based on its standard deviation, for the reproducibility error for $D_0$.

## Results and Discussion

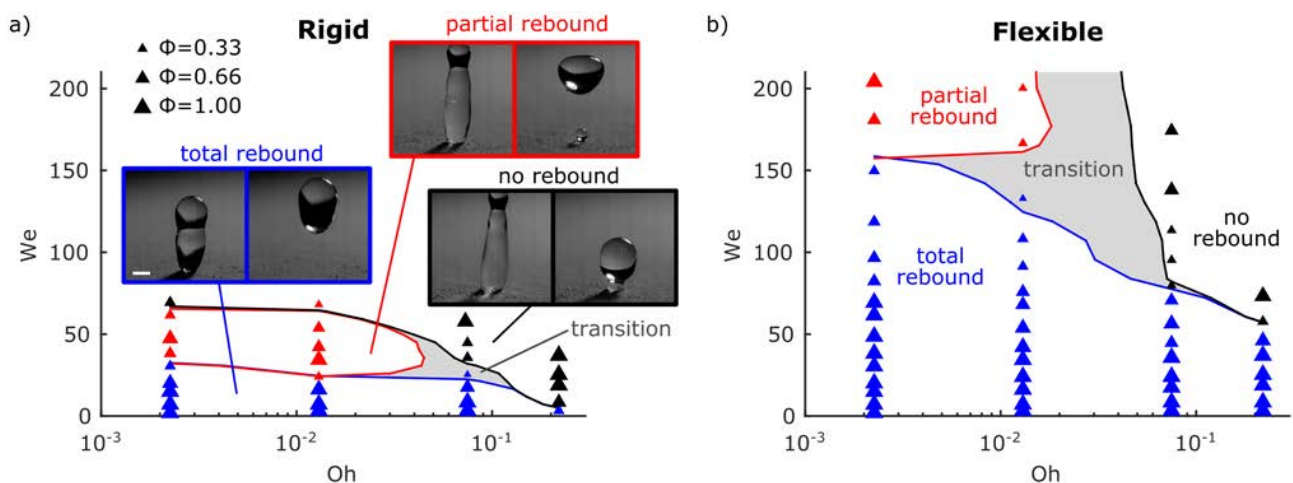

**Figure 1.** The effect of droplet viscosity on the rebound behavior of droplets impacting on flexible and rigid superhydrophobic substrates. Droplet-substrate impact outcome vs. $We$ and $Oh$ for a) rigid and b) flexible substrates ( $D_0 = 2.1 - 2.4$ mm). Symbols represent the highest probability of a given outcome: "total rebound" ( $\Phi_r = \max(\Phi_r, \Phi_p, \Phi_n)$ , ▲), "partial rebound" ( $\Phi_p = \max(\Phi_r, \Phi_p, \Phi_n)$ , ▲) and "no rebound" ( $\Phi_n = \max(\Phi_r, \Phi_p, \Phi_n)$ , ▲); the surface area of the triangles scales with $\Phi$ (see legend). In order to vary $Oh$ , various concentrations of glycerol in water were used for the droplets (0, 50, 75, and 85 wt. % glycerol stated with increasing $Oh$ ). The sample size of independent experiments ( $n$ ) is 5, and the number of total experiments (with replications, $n'$ ) is 15. The substrate was treated with nC1. Lines represent approximate transitions between different impact outcomes. Shaded gray areas labelled as "transition" indicate the regions where all 3 outcomes were observed but none of the associated probabilities is greater than 0.5. Insets images in a) show representative behavior for the different impact outcomes. Scale bar: a) 1mm.

*Droplet mobility – impalement resistance*

Figure 1 shows the probabilities of the different outcomes for droplet collisions on (a) rigid and (b) flexible substrates for a range of $We$ and $Oh$; the color of the triangles corresponds to the outcome with the higher probability and its surface area to the respectful value (examples for each outcome are shown as insets in Figure 1a). The lines mark the transition between the different regimes, which we draw using the following procedure. We assume that $\Phi_r$, $\Phi_p$ and $\Phi_n$ are smooth functions (surfaces) on the $We$ and $Oh$ plane. We used cubic smoothing splines to interpolate the surface values in the whole plane given our experimental data. The lines separating the different regions were calculated as the isolines for $\Phi_r$, $\Phi_p$ and $\Phi_n = 0.5$. There are regions on the $We$ and $Oh$ plane where we observed all of the possible outcomes and none of the associated probabilities exceeded the value of 0.5. We shaded these regions gray in Figure 1 and labelled them as "transition".

For the flexible case, high droplet mobility is sustained for a much wider region of $We$ and $Oh$. We observed an approximately 4-fold increase in the $We$ where the transition from total rebound to partial rebound or no rebound takes place. We attribute this enhancement in impalement resistance—the ability of the surface texture to resists meniscus penetration—to the substrate acceleration prior to contact with the droplet,[14] effect that is practically independent of $\mu$. We note though that in both the flexible and rigid cases, increasing $Oh$ reduced the value of $We$ where the transition from total-to-no rebound occurs since an increase in $\mu$ impedes recovering from partial penetration into the texture.[12] Additional information on characterizing the droplet impalement is given in SI 'Droplet Impalement' and Figure S3.

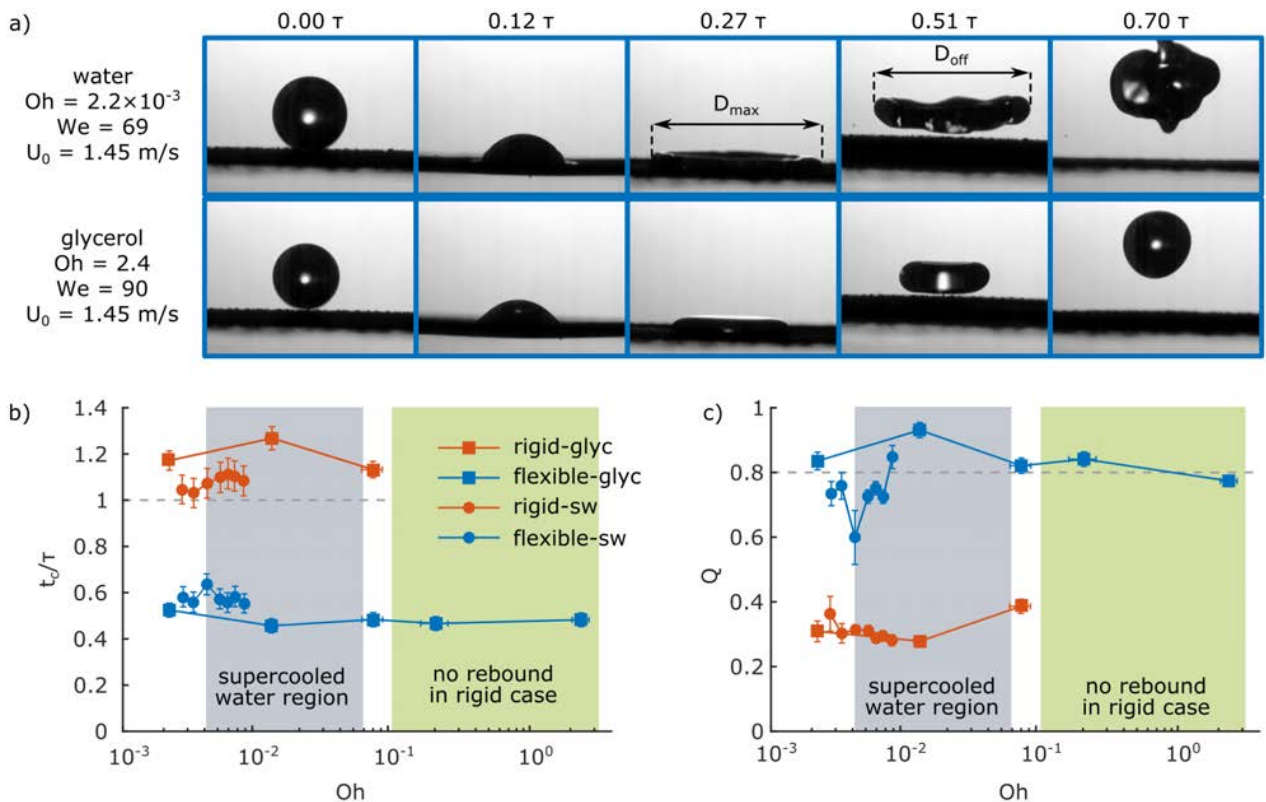

**Figure 2.** The effect of droplet viscosity on the contact time reduction for droplets impacting on a superhydrophobic substrate. a) Image sequence of water (top row; properties: $Oh = 2.2 \times 10^{-3}$, $We = 69$ and $U_0 = 1.45$ m·s⁻¹) and glycerol (bottom row; properties: $Oh = 2.4$, $We = 90$ and $U_0 = 1.45$ m·s⁻¹) droplets impacting on flexible substrate coated with nC2. The maximum spreading ($D_{max}$) and the take-off diameter ($D_{off}$) are also defined (top row). b) Dimensionless contact time ($t_c/\tau$) vs. $Oh$ for the rigid (■, ●) and flexible (■, ●) substrates ($\tau \approx 10$ ms, $D_0 = 2.2 - 2.4$ mm). $Oh$ was varied by either changing the supercooling of the water (●, ●, sw, $U_0 = 1.31$ m·s⁻¹, $n = 5$) or by the addition of glycerol (■, ■, glyc, $U_0 = 1.45$ m·s⁻¹, $n = 10$). The "gray" shaded area indicates the $Oh$ range that corresponds to supercooled water with a temperature range of −1 °C to homogeneous nucleation temperature[44] of −38 °C for $D_0 = 2.2$ mm. The "green" shaded area indicates the $Oh$ range where we observed no droplet rebound

from rigid substrates. The dashed gray line indicates the theoretical minimum[47] of $t_c/\tau$ for normal rebound of inviscid droplets impacting on a rigid substrate (single oscillation of a droplet). c) Pancake quality, $Q = D_{off}/D_{max}$, vs. $Oh$ for the same experiments as in panel b.

*Droplet mobility – droplet-substrate contact time reduction*

Additionally, droplet mobility can be enhanced by reducing the droplet-substrate contact time, $t_c$. Figure 2a shows an image sequence of water (top row) and glycerol (bottom row) droplets impacting on a coated (superhydrophobic, nC2) flexible substrate for similar values of $U_0$ and $\tau$ (see also Movie S1). In both cases, the droplets levitate and separate from the surface in a pancake shape, an indication of $t_c$ reduction.[*] Figure 2b plots $t_c/\tau$ vs. $Oh$ for supercooled water and water-glycerol droplets impacting on flexible and rigid substrates, where $\tau = \pi/4 \times \sqrt{\rho D_0^3/\sigma}$ is the inertial-capillary time. For liquid droplets with low viscosity, $\tau$ is equal to the theoretical period of natural oscillation of the droplet.[48] For the same substrate and coating type, we observed a good agreement between the water-glycerol and the supercooled water droplet impact experiments in terms of $t_c/\tau$. By using flexible substrates, one can achieve a 50 % reduction in $t_c$ compared to the rigid case. We note that for $Oh > 0.1$, droplets that impact rigid substrates were unable to rebound from the surface, contrasting the large $t_c$ enhancement resulting from substrate flexibility. Additionally, Figure 2c is a plot of the "pancake" quality,[*] $Q = D_{off}/D_{max}$, vs. $Oh$, where $D_{max}$ is the droplet diameter at maximum extension and $D_{off}$ is the diameter of the droplet when it separates from the surface. Previous work[*] suggests that $Q > 0.8$ is necessary for an impact event to be considered pancake bouncing, which we observe for water-glycerol droplet mixtures impacting on superhydrophobic flexible substrates.

Figure 3a presents the rebound behavior (reduced vs. normal $t_c$ modes, representative experiments are presented in Movie S2 for water and Movie S3 for glycerol) for a range of $We$ and $Oh$. It also plots $We_c$ vs. $Oh$ where the transition between the two rebound modes takes place. For $We < We_c$, the droplets rebound resembles the one observed for rigid surfaces with low viscosity liquids; the droplet retracts and used the stored surface energy to recoil with $t_c$ that is comparable to $\tau$. This is in accordance with previous reports for water droplet-substrate impacts on elastic superhydrophobic substrates.[35] For viscous droplets impacting on the flexible substrates, the rebound mode for $We < We_c$ resembles the one described before (water impacting rigid substrates, see Movie S2 and Movie **S3** for comparison), even though we did not observe rebound in the rigid case. In order to explore the limit of our flexible surfaces, we impacted an ultra-viscous and sticky liquid, namely honey (17 wt. % water content). The values $\rho = 1472$ kg·m[-3] and $\mu = 22.55$ Pa·s were taken from literature[40] and $\sigma = 76.8 \pm 1.8$ mN·mm[-1] was measured using the pendant droplet technique (see experimental section). For $D_0 \approx 2.8$ mm, we determined experimentally that $We_c = 18 \pm 10$ using the same procedure as before. In this case, the droplets impacting with $We < We_c$ were unable to rebound and remain stuck on the surface, whereas for $We > We_c$ we observed rebound (see Figure S5 and Movie S4). In the flowing we characterize the substrate behavior upon impact, and we employ a scaling analysis based on energy conservation in order to predict the onset of $t_c$ reduction and rationalize our results.

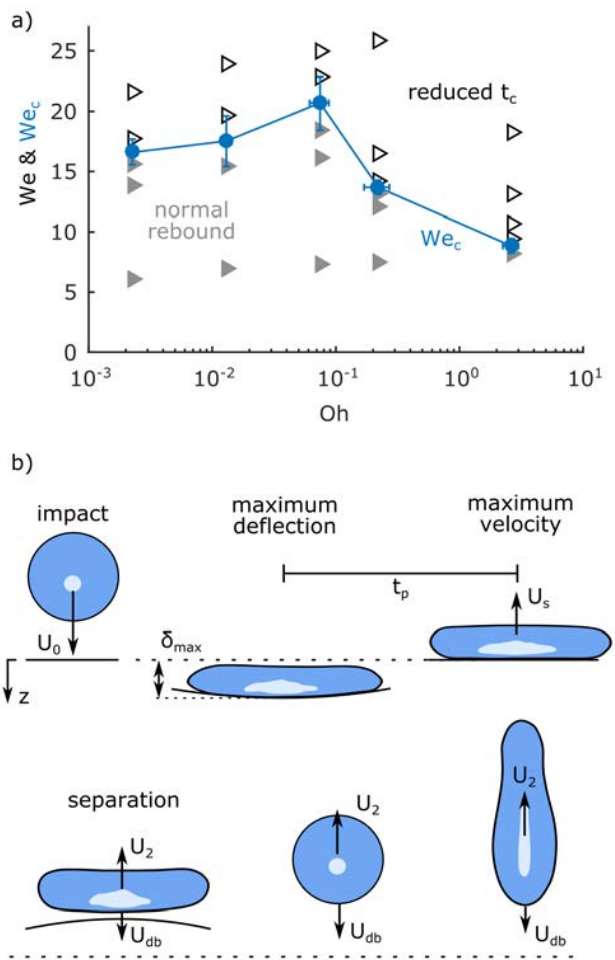

**Figure 3.** The mechanism of viscous droplet-substrate contact time reduction. a) Experimental rebound behavior vs. $We$ and $Oh$: reduced ($\triangleright$) vs. normal ($\blacktriangleright$) $t_c$ for $D_0 = 2.1 - 2.4$ mm. We also plotted is $We_c$ vs. $Oh$ ($-\bullet-$) with the vertical error bars denoting the span from the highest tested value where normal rebound was to detected to the lowest value which resulted in reduced $t_c$. Sample size ($n$) is 5 repetitions per point. b) Schematic showing the relevant stages during droplet impact and early rebound (just before impact, at substrate maximum deflection, at substrate maximum velocity, at the point of separation and oscillating airborne from left to right). The droplet has a velocity $U_s$ when it passes the point $z = 0$ for the 2nd time. After separation, the droplet center-of-mass is moving with velocity $U_2$ while its base has a relative velocity $U_{db}$.

*LDPE film energy rectification*

In order to quantify the role of the film on the droplet rebound, we extended the set of our experiments with droplet impact using $D_0 = 1.8$ to 2.9 mm for water and glycerol. We measured the maximum deflection of the substrate after impact, $\delta_{max}$, and the time, $t_p$ needed for the droplet-substrate system to travel from $\delta_{max}$ to its initial undisturbed position ($z = 0$, see Figure 3b for definition). We assume that the droplet-substrate system oscillates harmonically during this time, that is $\delta = \delta_{max}\cos(\omega_s t + d)$, where $\delta$ is the substrate deflection at time $t$, $\omega_s \approx \pi/(2t_p)$ is the angular velocity of the oscillation and $d$ is an initial phase. Here, we neglect the energy losses in the substrate for simplicity, but as we see later this assumption does not induce a serious error. As a result, the amplitude of the velocity during the oscillation is $U_s \approx \delta_{max}\pi/(2t_p)$ (see Figure 3b for a definition of $U_s$).

Figure 4a plots experimental values of $U_s$ vs. $U_0$ and exposes a linear trend between the two. It is then useful to introduce the rectification coefficient, $c_u = U_s/U_0$, which captures the portion of momentum the droplet has during the droplet-substrate oscillation in comparison to the one before impact. We estimated $c_u = 0.425$ by weighted linear least squares fitting (dashed line in Figure 4a). Notably, the mass of the impacting droplet has no effect on $U_s$. This behavior can be interpreted by idealizing the droplet-substrate system as a mass oscillating by two springs connected in series, one arising from the droplet deformability, $k_d$, and one from the film stiffness, $k_f$. Here, we assume that the film mass is negligible. If additionally the ratio of the film and droplet stiffness is constant, then the film will absorb and rectify a constant portion of the total energy of the system, that is $c_u^2 = k_d/(k_d + k_f)$. Hence, the film converts a fraction of the droplet kinetic energy ($E_{k0} = \pi D_0^3 \rho U_0^2 / 12$) as elastic strain energy (which scales with $\delta_{max}^2$) resulting in $\delta_{max} \propto \rho^{0.5} D_0^{1.5} U_0$. For the oscillating system, since the mass comes from the droplet, the angular velocity of the droplet-substrate is proportional as $\omega_s \propto \rho^{-0.5} D_0^{-1.5}$. By multiplying these terms, $U_s = \delta_{max} \omega_s \propto U_0$, the effect of droplet mass on $U_s$ vanishes and it is only dependent on $U_0$. In other words, heavier droplets will deflect the substrate more, but the oscillation will take place in at a lower frequency. In general, when the droplet mass is much greater than the effective mass of the substrate, $c_u$ is related to the relative stiffness between substrate and droplet. If the substrate mass is not negligible, then the ratio of the substrate to the droplet mass defines $\delta_{max}$ through an inelastic collision model,[50,51] and therefore $c_u$.

To further support this assumption, we measured the velocity of the droplet after separation, $U_2$, which is equal to $U_s$ reduced by the energy needed to separate the droplet from the surface. This energy is defined as the practical work of adhesion,

$$W_{ad} \approx \frac{\pi D_{max}^2}{4} \sigma \left(1 + \cos\theta_r\right) \tag{1}$$

We measured $U_2$ by tracking the droplet centroid in our image sequences and fit a parabolic trajectory. For low $Oh$, the droplets contacted the substrate for a second time after the initial separation, so we took care to estimate $U_2$ using only the frames in between. The kinetic energy of the droplet after separation is $E_{k2} = E_{k1} - W_{ad}$, where $E_{k1} \approx \pi D_0^3 \rho U_s^2 / 12 \propto c_u^2 U_0^2$. In terms of the coefficient of restitution, $\varepsilon = U_2 / U_0$, $\varepsilon^2 \approx c_u^2 - 3\xi^2 \left(1 + \cos\theta_r\right) / We$, where $\xi = D_{max} / D_0$ is the spreading parameter. Since for $Oh > 0.1$, the droplets do not rebound from the rigid substrate, we assume that independent of the outcome, either reduced $t_c$ or normal rebound, $U_2$ is attributed solely to the substrate action. For highly viscous fluids, $\xi \approx b_v Re^{0.2}$, where $b_v$ is the scaling factor between the $\xi$ and $Re^{0.2}$ (see SI 'Maximum spreading'); therefore:

$$\varepsilon = \sqrt{c_u^2 - \frac{3b_v^2 \left(1 + \cos\theta_r\right)}{We^{0.8} Oh^{0.4}}} \qquad (2)$$

We plot eq. (2) using the values for the glycerol ($\theta_r = 141°$ and $Oh = 2.6$) in Figure 4b, along with the experimentally measured values for $\varepsilon$. For better inspection, the values for fluids with $Oh > 0.1$ are plotted in color and the rest are shown in gray. The predicted curve captures the experimental trend while slightly overestimating $\varepsilon$, most likely due to the neglected energy losses on the film, but still the error is relatively small. In the case of $Oh < 0.1$, the droplet stores surface energy, which releases during its retraction. For $We$ lower or close to $We_c$, part of this energy assists the droplet in achieving higher $U_2$ which is reflected by the deviation of $\varepsilon$ towards higher values. As $t_c$ reduction takes place, the value of $\varepsilon$ converges for all $Oh$ to a constant value.

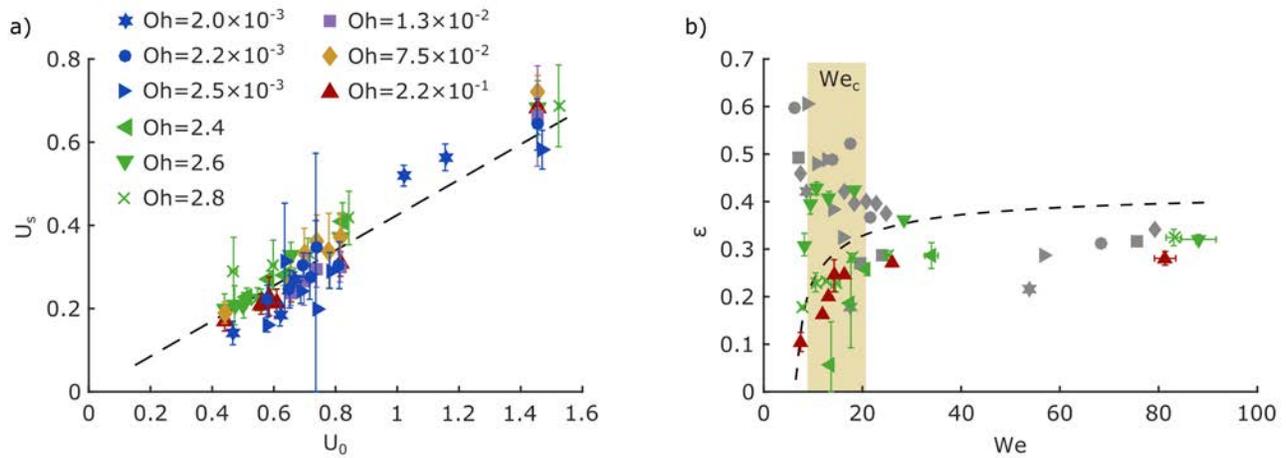

**Figure 4.** LDPE film ability to rectify energy. a) Droplet-substrate velocity after impact ($U_s$) vs. impact velocity ($U_0$) for different $Oh$ (water $D_0 = 2.9$ mm, water $D_0 = 2.4$ mm, water $D_0 = 1.9$ mm, 50 wt. % glycerol $D_0 = 2.2$ mm, 75 wt. % glycerol $D_0 = 2.1$ mm, 85 wt. % glycerol

$D_0 = 2.1$ mm, glycerol $D_0 = 2.6$ mm, glycerol $D_0 = 2.1$ mm, glycerol $D_0 = 1.8$ mm stated with increasing $Oh$ ). Dashed line represents the linear fit (slope 0.425) . Sample size ( $n$ ) is 5 repetitions per point. b) Coefficient of restitution ( $\varepsilon$ ) vs. $We$ for the same experiments as in panel a). Colored points correspond to mixtures with $Oh > 0.1$ that would not normally rebound from a rigid surface and gray points to mixtures with $Oh < 0.1$ . Markers as in panel a). The theoretical prediction (eq. (2)) is shown by the dashed (– –) line. The shaded area marks the $We$ region where the transition to pancaking occurs. Sample size ( $n$ ) is 5 repetitions per point.

*Energy balance for contact time reduction*

In our experiments, we observed that for highly viscous droplets, which are unable to recoil after impacting on rigid substrates, the rebound mode on flexible substrates shifts from reduced $t_c$ to normal with decreasing $We$ (see Movie S3). Accordingly, $E_{k2}$ in this case allows for droplet rebound but not for $t_c$ reduction. We hypothesize that as the droplet is close to the separation point, it still oscillates and such oscillations can hinder the droplet rebound in a pancake shape; therefore, one should eject the droplet fast enough such that the droplet oscillation is decoupled from the recoil dynamics (see Figure 3b). Hence, we are motivated to introduce a minimum of excess kinetic energy needed for $t_c$ reduction, which we denote in the following as $E_{k,ex}$ . Moreover, gravity does not play an important role in the process; for a rigid body in contact with an oscillating elastic substrate, separation will occur if the acceleration due to the oscillation will surpass the gravitational acceleration, $g$ , or stated mathematically $\omega_s U_s > g$ . Typical values for $\omega_s$ are in the order of 500 rad·s⁻¹, allowing for separation even at $U_s$ in the order of 0.01 m·s⁻¹. This value is well below the $U_2$ which we recorded on the onset of $t_c$ reduction in our experiments.

We speculate that $E_{k,ex}$ is connected to the retraction velocity of the droplet. The retraction of the droplet pushes its base towards the surface, counteracting the droplet-substrate separation. To estimate the $E_{k,ex}$, we assume that just before separation the base of the droplet is moving with a velocity of $U_2 - U_{db}$, where $U_{db}$ is the downward velocity of the base in a frame moving with the droplet (see Figure 3b). We expect $U_{db}$ to be proportional to retraction rate of the droplet which scales as $D_{max}/\tau_r$, where $\tau_r$ is equal to $\tau$ for inertial dominated retraction and to the viscous relaxation time, $\tau_v = \mu D_0 / \sigma$, for viscous dominated retraction.[52] The proposed crossover between the two regimes has been placed at $Oh = 0.05$,[52] but in our experiments values closer to $Oh \approx 1$ seem to be better a choice (for example $\tau_v$ is 3.5 times smaller than $\tau$ at $Oh \approx 0.22$ resulting in an overestimation of the retraction rate). Accordingly, in order to overcome the effect of droplet retraction $E_{k,ex} \sim \pi D_0^3 \rho U_{db}^2 / 12 \sim \pi D_0^3 \rho / 12 \times \left(D_{max}/\tau_r\right)^2$ and in a normalized form $E_{k,ex} / E_{k1} \sim \left(\xi\tau\right)^2 / \left(c_u^2 We\, \tau_r^2\right)$. In the same manner, we normalize the practical work of adhesion $W_{ad} / E_{k1} \approx 3\xi^2 \left(1 + \cos\theta_r\right) / \left(c_u^2 We\right)$. For any given fluid, we observe that the total energy resisting the $t_c$ reduction is $\left(W_{ad} + E_{k,ex}\right) / E_{k1} \sim \xi^2 / We$, which motivates the introduction of an early jumping parameter $P_j = \xi / \sqrt{We}$. Figure 5 plots the experimental values for $P_j$ at the onset of $t_c$ reduction vs. $Oh$. We define a critical value, $P_{j,c}$, in order to predict a type of rebound; for $P_j > P_{j,c}$, normal or no droplet rebound is expected whereas for $P_j < P_{j,c}$ reduced $t_c$ bouncing. The theoretical value for $P_{j,c}$, given the relations for $W_{ad}$ and $E_{k,ex}$, is

$$P_{j,c} \approx \frac{c_u}{\sqrt{3\left(1 + \cos\theta_r\right) + c_r \left(\tau/\tau_r\right)^2}} \tag{3}$$

where $c_r$ is the numerical coefficient connecting $E_{k,ex}$ to $\pi D_0^3 \rho / 12 \times \left( D_{max}/\tau_r \right)^2$. The calculation of $P_{j,c}$ takes into account the surface wettability, the ability of the substrate to rectify the elastic strain energy to the droplet and the liquid properties. We identify two limiting cases: a) at the low viscosity limit, e.g. for water, where $\xi \approx b_c We^{0.25}$ and $\tau/\tau_r = 1$ and b) at the high viscosity limit, e.g. glycerol, where $\xi \approx b_v Re^{0.2}$ and $\tau/\tau_r = \pi/\left( 4 Oh \right)$ (see SI 'Maximum spreading' for more on the spreading dynamics on flexible substrates), where $b_c$ and $b_v$ are proportionality coefficients.[53] At the low $Oh$ regime, the value of $P_{j,c}$ is independent of the liquid properties

$$P_{j,c} \approx \frac{c_u}{\sqrt{3\left(1+\cos\theta_r\right)+c_r}} \qquad (4)$$

and $We_c = \left( b_c / P_{j,c} \right)^4$. On the contrary, at the high $Oh$ regime,

$$P_{j,c} \approx \frac{c_u}{\sqrt{3\left(1+\cos\theta_r\right)+\left(\pi^2/16\right)c_r Oh^{-2}}} \qquad (5)$$

which results in lower values for $We_c$ with increasing $Oh$. We solve for the value of $We_c$, as:

$$We_c = Oh^{-0.5} \left( \frac{b_v}{c_u} \right)^{2.5} \left[ 3\left(1+\cos\theta_r\right) + \frac{\pi^2}{16} c_r Oh^{-2} \right]^{2.5} \qquad (6)$$

We plot the theoretical values for $P_{j,c}$ using the measured values for $\theta_r$ and $c_r = 0.51$ for the whole range of $Oh$ at Figure 5. The equation for viscous dominated retraction is employed above $Oh > 1$. Good agreement between the theoretical and the measured values is observed.

We also note that in the case of honey, we expect $E_{k,ex} = 0$ and thus $W_{ad}$ is the only factor impeding the rebound. Based on the measured $\theta_r = 142 \pm 10°$ and $\alpha = 3.6 \pm 1.7°$ for 12 μL honey droplets, we calculated a value for $W_{ad}$ based on the contact diameter 14 times lower than the one estimated by balancing $E_{k1}$ at $We_c$. This discrepancy most likely arises from the fact that calculation of $W_{ad}$ is not capturing the complex mechanism involved in the dewetting of such a liquid. The creation and disruption of capillary bridges on the texture asperities[54] and viscous dissipation[55] may increase $W_{ad}$. The non-Newtonian nature of honey[49] can also impede its dewetting.[56]

The reduction in $t_c$ and the repellency of highly viscous fluids by flexible substrates is based on the ability of the substrate to absorb $E_{k0}$ and rectify it back to the droplet aiding separation. Additionally, the timing of the energy rectification is an important factor; the droplet-substrate system should oscillate sufficiently fast for the droplet to recoil earlier in time. We detected separation for the droplet-substrate system at the highest point of its oscillation, namely at time $\sim 3\pi/(2\omega_s)$ after impact. Thus, the expected $t_c/\tau \approx 3\pi/(2\omega_s\tau)$, which results in ~0.5 for the typical values of $\omega_s$ and $\tau$ from our experiments and matches well with the observed one. For highly viscous liquids that do not recoil from the rigid surface, these timing considerations are irrelevant.

Contact time reduction on elastic substrates has been previously connected to splashing with gravity opposing its manifestation.[35] As we explained previously, gravity plays a minor role in the dynamics of the substrate-aided rebound. Moreover, the proposed Froude number criterion[35] for early droplet recoil, which connects the upward substrate inertia to gravity, $Fr = U_s/\sqrt{gD_{max}} > 1$, was valid in our experiments even for $We < We_c$, further supporting the that gravitational effect can be neglected. Most importantly, the previous model for predicting $t_c$ reduction relates its onset with splashing, a counter-intuitive connection. This implies that $We_c$ is independent of the substrate properties, like its ability to rectify energy, and depends solely on the surface roughness and liquid properties. Specifically, with increasing values of $\mu$, the $We_c$ should also increase in order for the splashing conditions to occur. In our experiments, we observe the opposite trend. In addition, for water, $We_c$ is notably lower than the previously reported value for the same $D_0 = 2.3$ mm (around 17 in this study compared to 51) and well away from the splashing regime. In summary, the suggested mechanism in this study differs to the previously reported one, since it connects the early droplet recoil mainly to the action of the substrate and to the surface wettability, and not to the action of the air layer under the leading edge of the spreading droplet.[35]

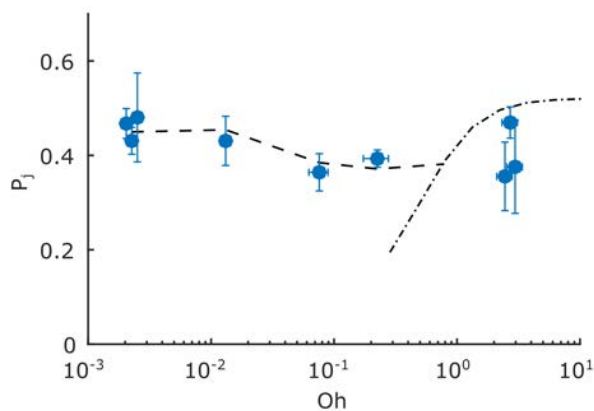

**Figure 5.** Jumping parameter, $P_j$, for the flexible substrates. $P_j$ at the onset of $t_c$ reduction vs. $Oh$ (same points denoted by $We_c$ in Figure 3a, including points with $D_0 = 1.9$ and 2.9 mm for water droplets and $D_0 = 1.8$ and 2.6 mm for glycerol). Low-viscosity (– –, eq. (4)) and high-viscosity (–.–, eq. (5)) theoretical values for the critical jumping parameter, $P_{j,c}$, below which reduced contact time is expected.

*Repellency of solidifying droplets*

The repellency of solid-like droplets and the ability of flexible surfaces to rectify the incoming kinetic energy to outgoing kinetic energy and overcome adhesion, motivate their use in repelling solidifying droplets. In the case of real life applications, solidification of supercooled water is more likely to occur in the presence of contamination on the surface rather than as a homogeneous process, a fact that is also observed in our experiments. Briefly, on the AgI contaminated surfaces, we always detected solidification of the impacting droplets, whereas for clean surfaces we detected solidification for a few impacts only at temperatures below −15 °C. On the other hand, the rate of solidification, $1/t_f$, where $t_f$ is the time elapsed from substrate contact until the completion of the first stage of freezing[10, 57] was significantly varied. We determined the completion of the freezing process optically, by the total clouding of the droplet volume and the standstill of the droplet surface (see Figure 6a and Movie S5 for definition). In accordance with theory, higher levels of subcooling and contamination, $\psi_c$, increased $1/t_f$. In order to quantify the probability of droplet solidification, we calculated the normalized rate of critical ice embryo formation, $J_n$, based on classical nucleation theory[23-26, 58] (see SI, 'Rate of critical ice embryo formation'). The higher the value of $J_n$, the more likely one is to observe the nucleation of the droplet and the faster the $1/t_f$.

Snapshots of a droplet impacting a highly contaminated surface are shown in Figure 6a. The first stage of solidification is completed fast enough, "recalescent freezing" which takes place in the order of ~10 ms,[10] resulting in the droplet remaining frozen almost at the point of maximum spreading. For the rigid case, rebound is completely suppressed as the excess surface energy remains stored in the partially frozen droplet and is not readily available after complete freezing. On the other hand, the droplet rebounds in the flexible case, due to efficient rectification of substrate strain energy back to outgoing droplet kinetic energy. Figure 6b shows the normalized rebound rate, $\tau/t_c$, and solidification rate, $\tau/t_f$, vs. $J_n$ ( $D_0 = 2.0 - 2.4$ mm, $We = 53 - 60$ and $T_d$ from −9 to −15 °C); $J_n$ was varied via droplet temperature and concentration of surface contamination, $\psi_c$. The absence of an event, either rebound or solidification corresponds to zero for the respective rate. When $\tau/t_f < \tau/t_c$ , rebound always occurred for droplets impacting both rigid and flexible substrates. This is because solidification takes place after droplet-substrate separation. If $\tau/t_f \approx \tau/t_c$, for impacts on the rigid substrates, the droplets do not rebound—the stored surface energy is not rectified back to kinetic energy, resulting in droplet arrest. The positive effect of increased $\tau/t_c$ on droplet rebound under solidifying conditions is in accordance with previous studies with macro-textured surfaces the molten metal droplets.[17] With a further increase in $\tau/t_f$, one would expect that the droplets would also stop rebounding from the flexible substrates as $\tau/t_f > \tau/t_c$, but this is not the case. The stored strain energy was able to overcome the water-ice slurry adhesion and the solidifying droplets were thrown away by the surface action (examples of the collision outcome for flexible and rigid substrates for varying $\tau/t_f$ are presented in Movie S6). We expect that the work of adhesion for water-ice slurry should be greater than $W_{ad}$, but to our knowledge, no analytical description is available. However, we anticipate that the surface wettability should play an important role since it has been shown that $W_{ad}$ is proportional to the strength of the ice adhesion.[25] Separately, high values of $c_u$, which can be tuned by the substrate stiffness, are also desirable in this case, since more energy will be available for shedding the ice-water mixture from the surface.

The repellency of solidifying droplets has a triple role in mitigating ice accretion. Firstly, flexibility is acting as both a mean for reducing $t_c$, lowering the probability of droplet arrest due to solidification, and as a passive ice shedding mechanism. Furthermore, the self-cleaning mechanism of superhydrophobic surfaces,[59] which is well understood in room temperature, is extended at the deeply supercooled range. The impacting droplets carry away the contaminating particles that promote ice nucleation thus lowering the readiness of the subsequent droplet to solidify. This can be a thought as a new mechanism that promotes icephobicity.

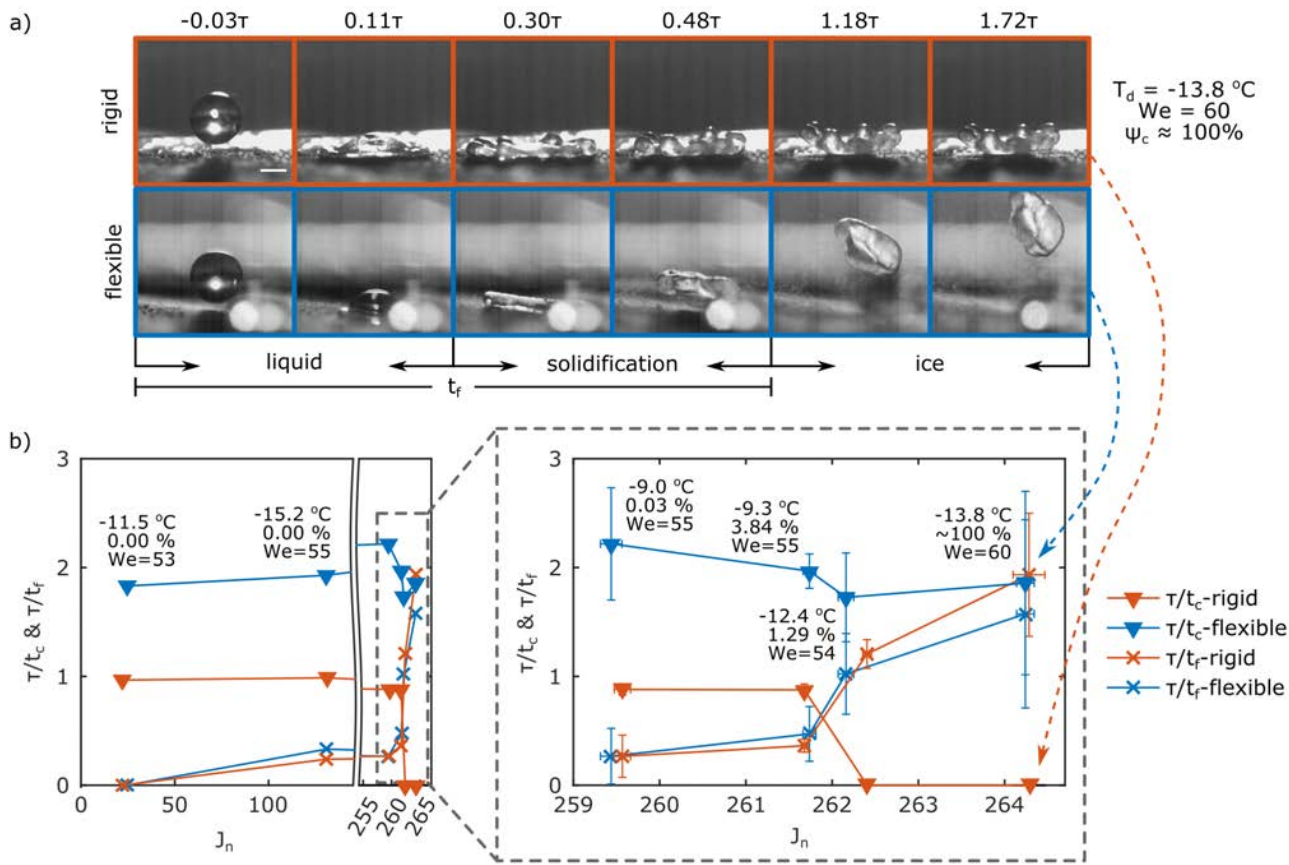

**Figure 6.** The role of flexibility in repelling partially solidified droplets. a) Snapshot of a supercooled droplet impacting on flexible and rigid substrates contaminated with AgI particles ( $\psi_c \approx 100\%$, $D_0 = 2.0$ mm, $We = 60$, $T_d = -13.8$ °C). b) Rebound ( $\tau/t_c$ ) and recalescence solidification rates ( $\tau/t_f$ ) vs. the normalized nucleation rate $J_n$ for rigid and flexible substrates ( $D_0 = 2.0 - 2.4$ mm, $We = 53 - 60$, $\psi_c = 0 - 100\%$ and $T_d$ from −9 to −15 °C). Also enlarged is the region where $\tau/t_f \approx \tau/t_c$. Rates with zero value indicate an absence of the respective event, either no rebound or no solidification. Next to each point for the flexible case we indicate the $T_d$, $\psi_c$ and $We$, from top to bottom respectively. For the rigid case, the impact and contamination conditions are similar to the ones reported for the flexible case for similar values of $J_n$. Error

bars are omitted in the left panel for better inspection. Sample size ($n$) is 4-5 repetitions per point.

## Conclusion

In this work, we reported the use of substrate flexibility, working collaboratively with surface micro/nanotexture, as a passive mean to repel supercooled and viscous droplets. We quantified these findings by the decline in the impalement probability, the reduction in the droplet-substrate contact time, and in the raise in droplet recoil velocity. We experimentally demonstrated these results for a range of droplet supercooling and viscosities and explained the underlying mechanisms with appropriate modeling (impalement: kinetic-capillary; substrate energy rectification and contact time reduction: kinetic-elastic-capillary). We showed that the repellency of highly viscous liquids, which do not recoil from rigid superhydrophobic substrates, is accomplished because of the efficient absorption and rectification of the kinetic energy of the droplet by the flexible substrate. Additionally, for low viscosity liquids, if the droplet-substrate system oscillation period is faster than the inertia-capillary time of the droplet, the upward movement of the substrate is decoupled from the droplet oscillation resulting in the observed contact time reduction. Further, we tested the flexible materials under challenging icing conditions and we presented supercooled droplet-substrate impact experiments on surfaces contaminated with ice nucleation promoters (AgI), where the time to partial droplet solidification was faster than the natural oscillation of a droplet, even at not extreme supercooling. This way, we demonstrated that flexible materials can even shed partially solidified droplets and overcome substrate adhesion using the abovementioned mechanism. Therefore, we conclude that the design of flexible icephobic materials should be based on their ability to store the droplet kinetic energy and release it a later time, facilitated by the low areal density and moderate stiffness, as our theoretical modeling reveals.

## Associated Content

**Supporting Information.** The following files are available free of charge.

Additional data and supporting figures on the surface impalement resistance characterization and on the droplet maximum spreading. Details on the calculation of the rate of the critical ice embryo formation. Supporting figures of the surface characterization, the experimental setup and the impacts of honey droplets. (PDF)

**Movie S1.** Contact time reduction on flexible surfaces. Side-by-side comparison of droplets of different viscosities (water and glycerol) and droplets of different diameter (water, $D_0 = 1.9$ and 2.9 mm) impacting on flexible superhydrophobic substrates. The substrates for all cases are LDPE films coated with nC2 and mounted under ~0.5 % strain. (AVI)

**Movie S2.** Critical $We$ for contact time reduction for water droples Side-by-side comparison of water droplets impacting on flexible superhydrophobic substrates above and below $We_c$. The rebound changes from normal to reduced $t_c$. Impacts of water droplet for the same $We$ as in the flexible case are included for comparison. The movie slows down close to the instance of droplet-substrate separation for better inspection. The substrates are LDPE films coated with nC2 and mounted under ~0.5 % strain. (AVI)

**Movie S3.** Critical $We$ for contact time reduction for glycerol droplets. Side-by-side comparison of glycerol droplets impacting on flexible superhydrophobic substrates above and below $We_c$.

The rebound changes from normal to reduced $t_c$. The substrates are LDPE films coated with nC2 and mounted under ~0.5 % strain. (AVI)

**Movie S4.** Honey droplets impact flexible superhydrophobic substrates. The honey droplets stick on the surface or rebound, depending on the $We$ of the impact. The substrates are LDPE films coated with nC2 and mounted under ~0.5 % strain. (AVI)

**Movie S5.** Measurement of the solidification rate ($1/t_f$) of supercooled droplets. Example of a droplet impacting a flexible surface contaminated with AgI particles. We define $t_f$ as the time elapsed from the instance of droplet contact with the nucleation promoting seed until the end of the recalescent freezing, indicated by the clouding of the whole droplet volume. The substrate is LDPE film coated with nC2 and mounted under ~0.5 % strain. (AVI)

**Movie S6.** Repellency of solidified droplets. Side-by-side comparison between the rigid and the flexible substrates for 3 cases with respect to the solidification ($1/t_f$) and the rebound ($1/t_c$) rates: a) $1/t_f < 1/t_c$ for both flexible and rigid substrates, b) $1/t_f \approx 1/t_c$ for the rigid case but $1/t_f < 1/t_c$ for the flexible case and c) $1/t_f > 1/t_c$ for both substrates. The substrates are LDPE films coated with nC2 and mounted under ~0.5 % strain. (AVI)

## Author information

### Corresponding author

E-mail : dimos.poulikakos@ethz.ch

E-mail : thomschu@ethz.ch

## Author Contributions

D.P., T.M.S and T.V. conceived and designed the research. T.V conducted the experiments and analyzed data. All authors wrote the manuscript.

**Notes**

The authors declare no competing financial interest.

**Acknowledgements**

Partial support of the Swiss National Science Foundation under grant number 162565 and the European Research Council under Advanced Grant 669908 (INTICE) is acknowledged.

## Graphical TOC

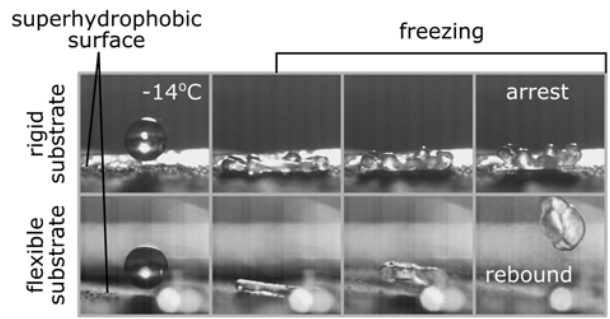

# Supplementary information

# "Imparting icephobicity with substrate flexibility"

*Thomas Vasileiou, Thomas M. Schutzius\*, Dimos Poulikakos\*.*

Laboratory of Thermodynamics in Emerging Technologies, Department of Mechanical and

Process Engineering, ETH Zurich, Sonneggstrasse 3, CH-8092 Zurich, Switzerland.

\* To whom correspondence should be addressed.

Professor Dimos Poulikakos
Laboratory of Thermodynamics in Emerging Technologies
Department of Mechanical and Process Engineering
ETH Zurich, ML J 36
8092 Zurich
Switzerland
Phone: +41 44 632 2738
Fax: +41 44 632 1176
Email: dpoulikakos@ethz.ch

Dr. Thomas M. Schutzius
ETH Zurich
Laboratory of Thermodynamics in Emerging Technologies
Sonneggstrasse 3, ML J 38
CH-8092 Zürich
SWITZERLAND
Phone: +41 44 632 46 04
Email: thomschu@ethz.ch

## Droplet Impalement

In addition to the droplet impact outcome map, one can also characterize the degree of droplet impalement with the recoil rate, $1/t_c$, and the impalement diameter, $D_{imp}$. In the absence of impalement, $1/t_c$ is almost constant (for a range of $We$ at a fixed value of $Oh$) and starts to decrease at the onset of impalement.[1,2] For partial impalement or no rebound we assume $1/t_c \approx 0$. In Figure S3, we plot $\tau/t_c$ and $D_{imp}/D_0$ vs. $We$ for the same experiments as in Figure 1, where $\tau = \pi/4 \times \sqrt{\rho D_0^3/\sigma}$ is the inertial-capillary time. As expected, $\tau/t_c$ starts to reduce at lower $We$ for the rigid case (slower rebound dynamics), which is attributed to fluid partially penetrating the surface asperities even in cases when the droplets eventually rebound. Likewise, when impalement did occur, $D_{imp}/D_0$ was always higher in the rigid compared to the flexible case for the same $We$ and $Oh$.

## Maximum spreading

We measured the spreading parameter $\xi = D_{max}/D_0$ from the droplet impacts on the flexible substrate (LDPE film coated with nC2) under 0.5 % strain. The response time of the substrate (time elapsed from impact to maximum deflection) is comparable to the timescale of the droplet spreading dynamics so we expect no detectable difference in $\xi$ in comparison to the rigid case.[3] For impacts on the rigid hydrophobic substrates, two scaling laws have been proposed depending on the dominant force resisting the droplet spreading, which can be either capillarity or viscosity: $\xi \sim We^{0.25}$ and $\xi \sim Re^{0.2}$ for low and high $Oh$ respectively.[4] Figure S4a and b shows $\xi$ vs. $We$ and $Re$; we included the points from all experiments but we show in color only the ones relevant for each scaling law. The theoretical predictions are plotted by the dashed lines. The proportionality coefficient for low $Oh$ case is $b_c = 0.9$, taken from literature, and for the high $Oh$ is $b_v = 1.33$, which we fitted in our data. Good agreement can be observed between theory and experiment.

In order to verify the use of the correct scaling law, we calculated the impact parameter, $P = We \, Re^{0.8}$ and we plot $\xi \, Re^{-0.2}$ vs. $P$ in Figure S4c. Values for $P < 1$ indicate capillary dominated spreading whereas for $P > 1$ viscosity dominated spreading. The reported values confirm the proper use of theory. Most importantly, we are able to use the theory that has been developed for rigid substrates in order to predict $\xi$ for impacts on flexible substrates.

## Rate of critical ice embryo formation

According to classical nucleation theory, the water molecules at the supercooled state create ordered clusters, called ice embryos.[5] When the embryos reach a critical size, $r_c$, are able to self-sustain and grow, provoking the freezing of the whole liquid phase. We can calculate $r_c = 2\gamma_{IW}/\Delta G_{f,v}$, where $\gamma_{IW}$ is the ice-water surface energy, $\Delta G_{f,v}$ is the volumetric free energy difference between bulk water and ice. The rate of critical ice embryo formation[6,7] is determined as

$$J = K(T)N_A \psi_c A(t) \exp\left(\frac{-\Delta G(T, \theta_{IW}, R)}{k_B T}\right) \qquad (1)$$

where $K(T)$ is a kinetic constant, $N_A$ is the areal density of water molecules in the liquid phase, $A(t)$ is the time-dependent droplet-substrate surface area, $k_B$ is the Boltzmann's constant, $T$ is the temperature and $t$ the time. $\Delta G(T, \theta_{IW}, R)$ denotes the free energy barrier of a critical size ice embryo formation, calculated as:

$$\Delta G(T, \theta_{IW}, R) = \frac{16\pi\gamma_{IW}^3}{3\Delta G_{f,v}^2} f(\theta_{IW}, R) \tag{2}$$

where $\theta_{IW}$ is the contact angle forming on the water-ice-surface triple line, $R$ is the surface roughness and $f(\theta_{IW}, R)$ is the factor relating of the energy barrier for homogeneous nucleation to the one in the heterogeneous case. For $R \gg r_c$, like in our experiments, this factor is predominantly determined by $\theta_{IW}$ as $f \approx (2 + \cos\theta_{IW})(1 - \cos\theta_{IW})^2/4$. Finally, the kinetic term is given as

$$K(T) = n_s Z \frac{k_B T}{h} \exp\left(\frac{-\Delta F_{diff}(T)}{k_B T}\right) \tag{3}$$

where $Z$ is the Zeldovich factor accounting for the depletion of ice embryos due to their production, $\Delta F_{diff}(T)$ is the diffusion activation energy, $h$ is the Planck constant and $n_s$ is the number of molecules in the embryo water interface. By combining equations (1) and (3), we can rewrite $J$ as

$$J = C_{pref} T \psi_c A(t) \exp\left(\frac{-\Delta F_{diff}}{k_B T}\right) \exp\left(\frac{-\Delta G}{k_B T}\right) \tag{4}$$

The prefactor $C_{pref}$ combines all the terms that remain almost constant with temperature. Furthermore, we introduce $J_n = \log_{10}(J/J_0)$ where we normalized $J$ by $J_0$, the rate of critical ice embryo formation for a droplet impacting an uncontaminated surface at $T_{ref} = -11\ °C$,

$$J_n = \log_{10}\left(\frac{T}{T_{ref}}\right) + \log_{10}(\psi_c) + \frac{1}{\ln(10)}\left(-\frac{\Delta F_{diff} + \Delta G}{k_B T} + \frac{\Delta F_{diff,ref} + \Delta G_{ref}}{k_B T_{ref}}\right) \tag{5}$$

In the last equation, we assume that $A(t)$ is almost identical for droplets having roughly the same $D_0$ and $We$. Moreover, for the contaminated surface case, we assume that the main contribution in $J_n$ comes from the AgI particles and we neglect any effect from the $1 - \psi_c$ part of the surface in the forming ice embryos.

In an overview of different formulation for the kinetic and thermodynamic parameters from literature, Ickes *et al.*[7] propose that the combination of $\Delta F_{diff}$ and $\gamma_{IW}$ as derived by Zobrist *et al.*[8] and Reinhardt and Doye[9] respectively leads to the best agreement with a dataset of homogeneous nucleation rates containing measurements from 33 freezing experiments. In our calculation, we used these recommended formulations. For the $\Delta G_{f,v}$, we utilized the free energies equations by Holten *et al.*[10] and Feistel and Wagner[11] for supercooled water and ice respectively, with equal values at the triple point as proposed by the International Association for the Properties of Water and Steam. For AgI particles $\theta_{IW} \approx 25°$ has been reported[12] and for the uncontaminated surface case we estimated[7] $\cos\theta_{IW} = \cos\theta(\gamma_I - \gamma_w)/\gamma_{IW}$, where $\gamma_I \approx 106$ mJ/m² is the ice-water surface energy.

**Supplementary figures**

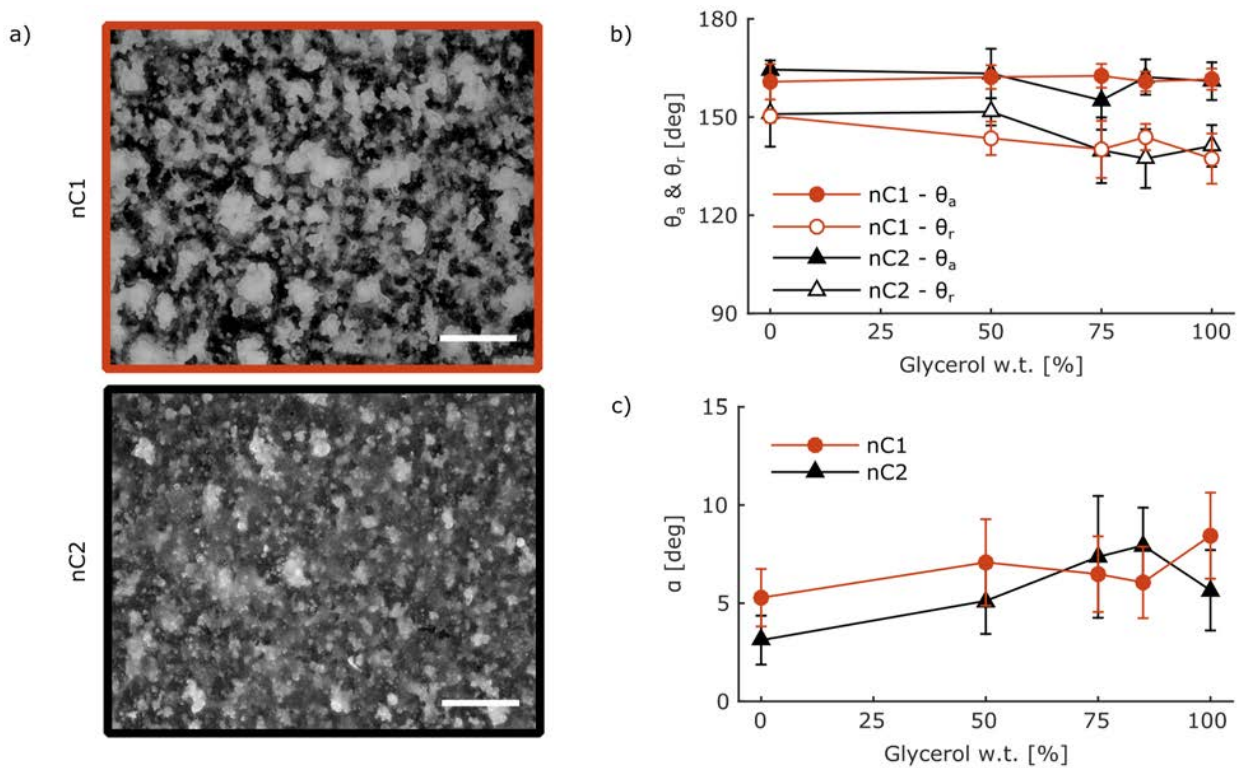

**Figure S1.** Characterization of superhydrophobic coatings. a) Micrographs of the surface produced by spray coating with nC1 (upper panel) and nC2 (lower panel) nanocomposites. b) Advancing ($\theta_a$, closed symbols) and receding ($\theta_r$, open symbols) contact angles of the two coating (circles for nC1, triangles for nC2) for varying concentrations of water-glycerol mixtures. Sample size ($n$) is 5 repetitions per point. c) Roll-off angle ($\alpha$) of the two coating (circles for nC1, triangles for nC2) for a 6 µL droplet at different water-glycerol mixture concentrations. Sample size ($n$) is 5 repetitions per point. Scale bar: a) 100 µm.

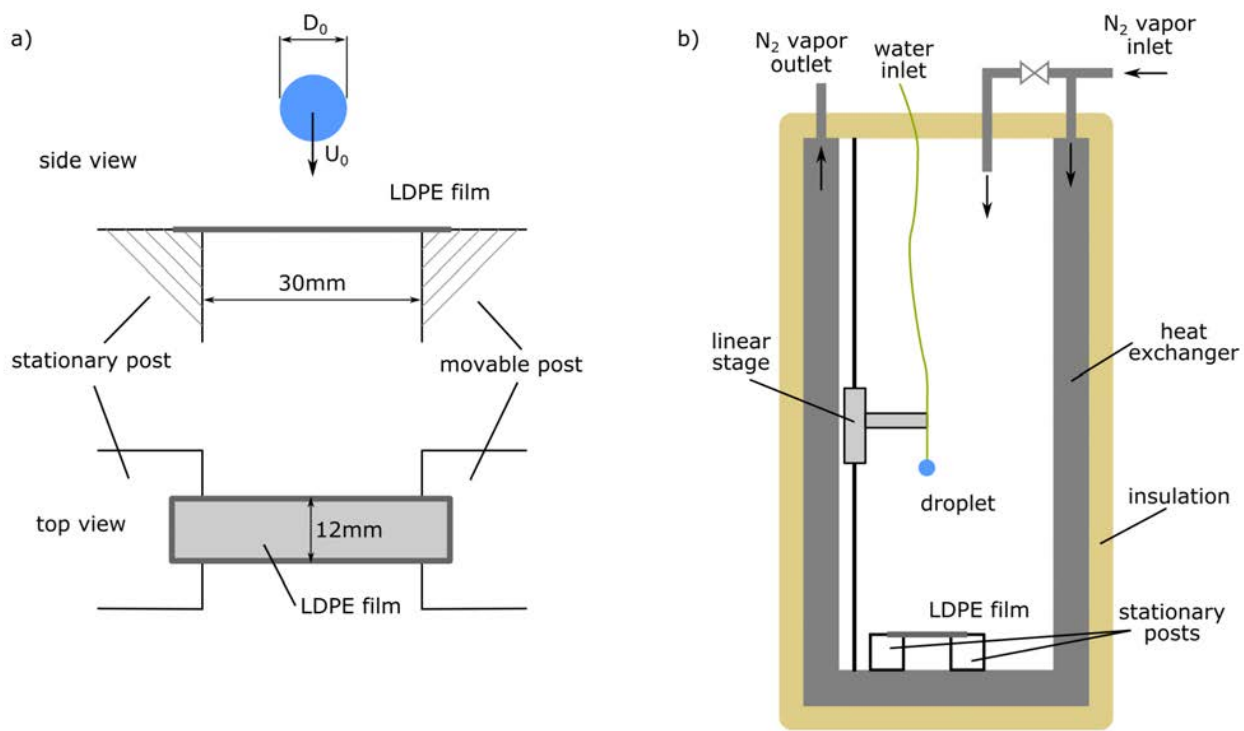

**Figure S2.** Droplet impact setup schematic. a) Side and top view of the LDPE film mounting for the ambient temperature experiments. The film is sectioned in 30 mm × 12 mm and is placed on one fixed and one movable posts. A strain can be applied to the film if desired, by displacing the movable post. b) Insulated chamber for low-temperature experiments. The temperature of the chamber is controlled by a $N_2$ vapor flow and the relative humidity is ~0 %. The droplet release height can be altered by a linear stage. In this case, the strain on the LDPE film is applied manually due to space restrictions.

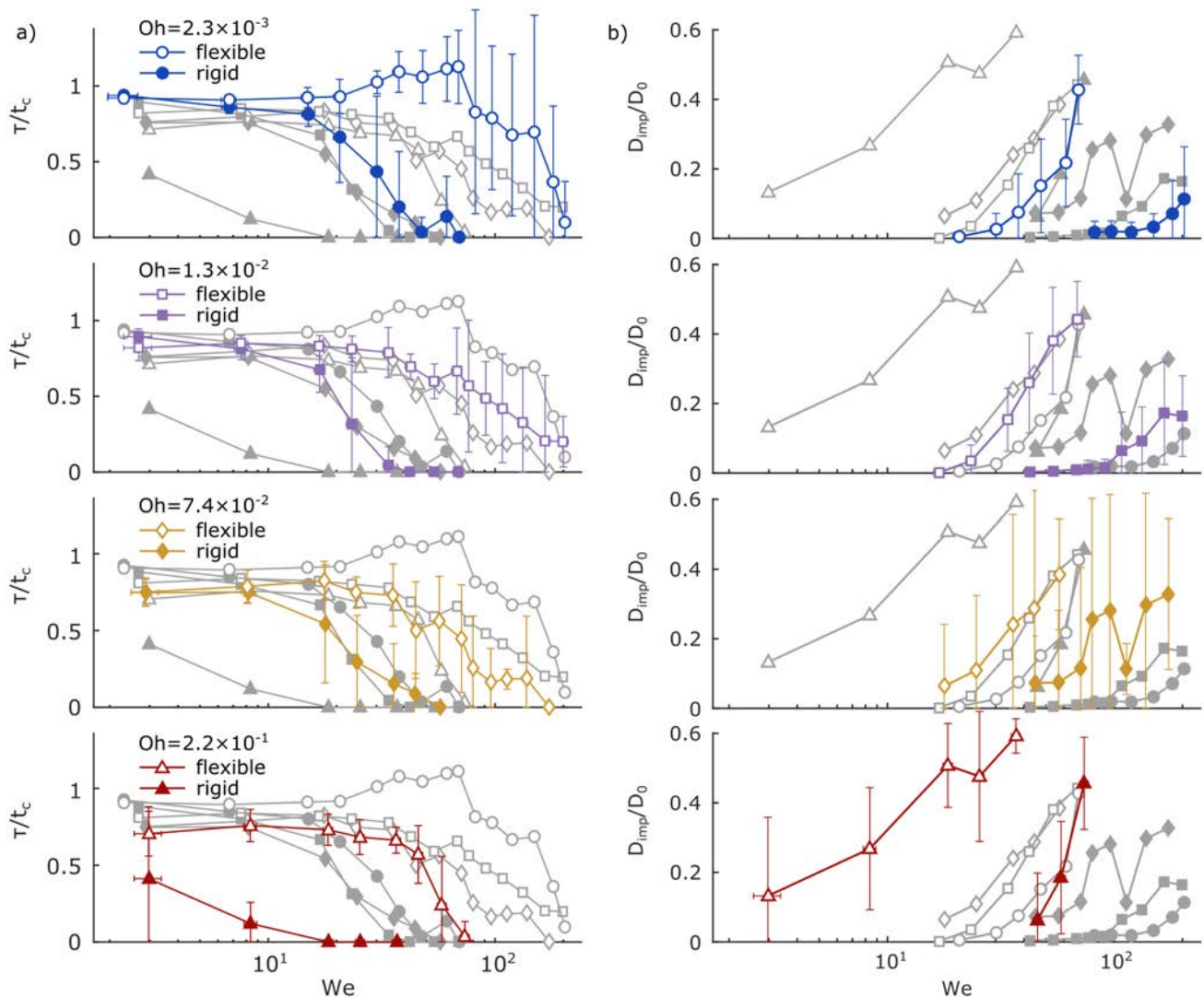

**Figure S3.** Comparing the contact time, $t_c$, and impalement diameter, $D_{imp}$, for droplets impacting on flexible and rigid substrates. a) Inverse of normalized contact time ($\tau/t_c$) vs. $We$ for impacts on flexible and rigid substrates treated with the hydrophobic coating nC1. Each panel corresponds to a different water-glycerol mixture (pure water and 50, 75 and 85 wt. % glycerol-water mixtures stated in increasing $Oh$ order, $n = 5$, $n' = 15$). To facilitate the comparison between flexible and rigid cases at each $Oh$, we plot the data from the remaining $Oh$ in gray without error bars in each panel. We set $\tau/t_c = 0$ for partial or no rebound. Experiments are the same as in Figure 1. b) Normalized impalement diameter ($D_{imp}/D_0$) vs. $We$. Same data

presentation as in panel a. Points with $D_{imp}/D_0 = 0$ have been omitted for better inspection.

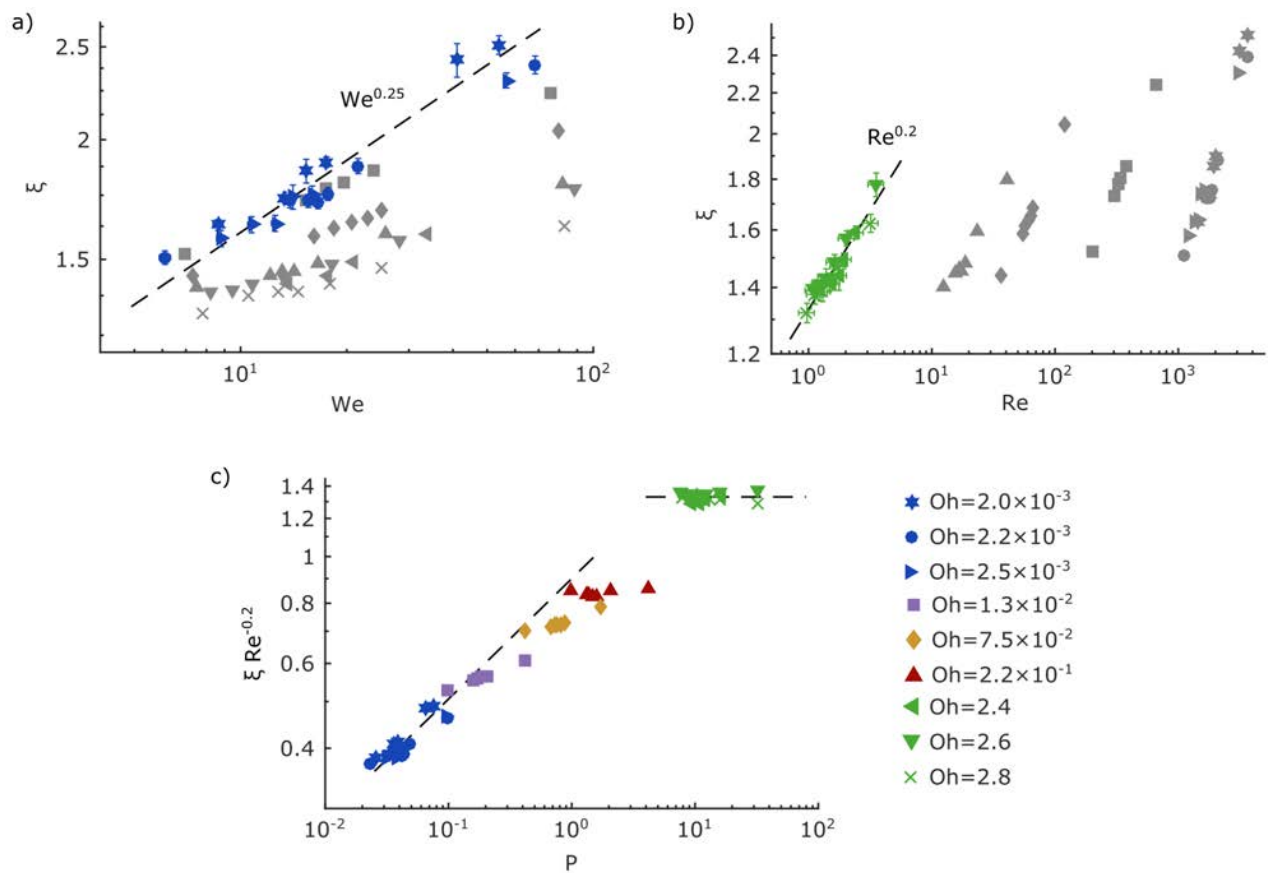

**Figure S4**. Maximum spreading of droplet impacting flexible substrates. a) Spreading coefficient

($\xi$) vs. $We$ for droplet impacting flexible substrate (LDPE film coated with nC2) under 0.5 %

strain. Markers in color correspond to experiments valid for the scaling law for inertia dominated

spreading ($\xi \sim We^{0.25}$ shown by the dashed (– –) line). Markers shown in gray represent

experiments that deviate from this theory and the error bars are omitted from these points for the

sake of clarity. Marker legend as in panel c. Sample size ($n$) is 5 repetitions per point. b)

Spreading coefficient ($\xi$) vs. $Re$ for the same experiments as in panel a. Markers in color correspond to experiments valid for the viscous dominated spreading ($\xi \sim Re^{0.2}$ shown by the dashed (– –) line). Markers shown in gray represent experiments that deviate from this theory and the error bars are omitted from these points for the sake of clarity. Marker legend as in panel c. Sample size ($n$) is 5 repetitions per point. c) Scaled maximum spreading ($\xi Re^{-0.2}$) vs. the impact parameter ($P$). The dashed lines indicate the scaling of the two regimes with the transition taking place around $P \approx 1$.[4]

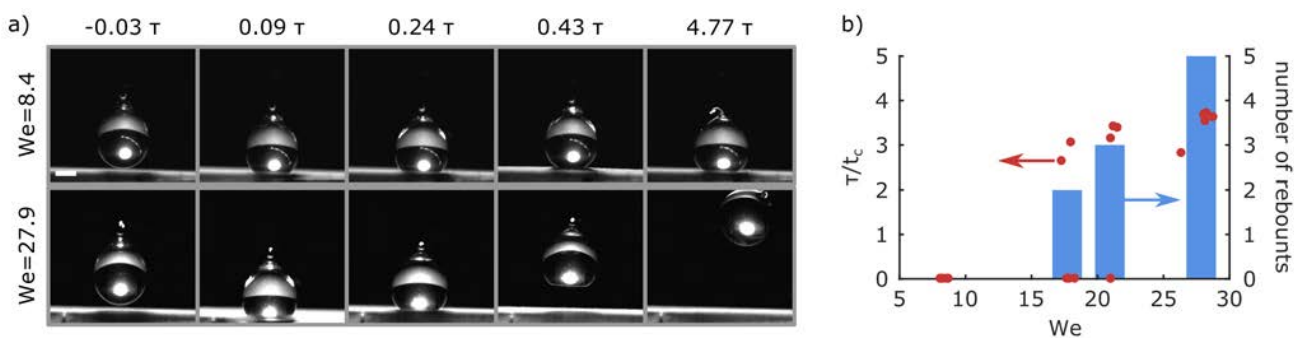

**Figure S5.** Repellency of honey droplets. a) Image sequence of honey droplets ($D_0 \approx 2.8\,\text{mm}$, $\tau \approx 16.1$ ms) impacting a flexible substrate coated with nC2. The upper snapshot sequence corresponds to $We = 8.4$ which results in the droplet sticking on the surface, whereas the lower sequence for $We = 27.9$ shows the honey droplet rebounding from the surface. b) Normalized reciprocal of the contact time ($\tau/t_c$, red circles corresponding to left axis) and the number of rebound (bar plot corresponding to right axis) vs. $We$ for the experiments performed to determine $We_c$ for the honey droplets. For no rebound $\tau/t_c = 0$. For the range of $We$ between 15

and 25, we observed both outcomes, droplets could recoil or stick onto the surface. Sample size (

*n* ) is 5 repetitions. Scale bar: a) 1 mm.